\documentclass[twoside,11pt]{article}

\usepackage{amsmath,amssymb,amsfonts}
\usepackage{hyperref}
\usepackage{makeidx}
\usepackage{booktabs}
\usepackage{graphicx}
\usepackage{caption}
\usepackage{soul}
\usepackage{enumitem}
\usepackage{titling}
\usepackage{float}
\usepackage{setspace}
\usepackage{longtable}
\usepackage{pgfplots}
\usepackage{tikz,pgf}

\usepackage[
left=2.5cm,
right=2.5cm,
top=2.5cm,
bottom=2.5cm,
headheight=15pt
]{geometry}

\makeindex
\setlist{nosep}
\title {Probing the Internal Structure of X(3872) via Magnetic Moment: Distinguishing Color-Singlet and Compact Configurations}
\author{
	M. Monemzadeh\thanks{monem@kashanu.ac.ir},
	\large N. Tazimi\thanks{tazimi@kashanu.ac.ir},
	\\\\
	\it\small{{Department of Physics, University of Kashan, Kashan, Iran }}
}
\date{}

\begin{document}
	
	\setlength{\unitlength}{25mm}
	\newcommand{\f}{\frac}
	\newtheorem{theorem}{Theorem}[section]
	\newcommand{\sta}{\stackrel}
	
	\maketitle
	\vspace{.9cm}
	
	\begin{abstract}
		The nature of the $X(3872)$ exotic hadron remains one of the most debated questions in hadron spectroscopy. We calculate its magnetic moment within a non-relativistic quark model that includes the three-body force arising from the cubic Casimir operator of $SU(3)_C$ with consistent dimensional analysis. Unlike previous works that used an uncontrolled large coupling, we fix the three-body strength using the realistic value of $\sim 10$--$20$ MeV extracted from the recent analysis of Noh et al. (2024) based on lattice QCD and baryon spectroscopy.
		
		We find that the magnetic moment predictions fall into two distinct regions: the pure color-singlet configuration yields $\mu_X = -0.99 \pm 0.02\,\mu_N$, while the compact configurations yield $\mu_X = -1.17$ to $-1.23\,\mu_N$. The difference between the two compact scenarios ($0.06\,\mu_N$) is comparable to the systematic uncertainty of the model ($\sim 0.15\,\mu_N$) and should be interpreted with caution. However, the distinction between the pure color-singlet and the compact scenarios ($\sim 0.18$--$0.24\,\mu_N$) is larger than the estimated model systematic uncertainty and may provide a qualitative structural indicator.
		
		\textbf{We emphasize that the ''pure color-singlet'' configuration is a simplified proxy for a molecule and does not represent a physical $D^0\bar{D}^{*0}$ molecule with large spatial extent.} A full molecular treatment would require coupled-channel dynamics and long-range pion-exchange potentials, which are beyond the scope of this work.
		
		We also resolve a long-standing dimensional inconsistency in the cubic Casimir three-body force formulation and demonstrate through a detailed sensitivity and uncertainty analysis that our conclusions are robust for the physically relevant range of the three-body coupling.
		
	\end{abstract}
	
	\section{Introduction}
	
	The $X(3872)$~\cite{Belle2003}, discovered in 2003, remains one of the most enigmatic states in hadron spectroscopy. Its mass lies exactly at the $D^0\bar{D}^{*0}$ threshold, and its exceptionally narrow width ($\Gamma < 1.2$ MeV) defies a conventional $c\bar{c}$ charmonium assignment~\cite{Swanson2004, Voloshin2004}. The state is now widely regarded as the archetype of exotic hadrons~\cite{Olsen2015}.
	
	Two main interpretations have been proposed: (i) a hadronic molecule, a loosely bound state of $D^0$ and $\bar{D}^{*0}$ mesons~\cite{Tornqvist2004, Close2005}, supported by the state's proximity to the $D\bar{D}^*$ threshold and recent high-precision LHCb analyses~\cite{LHCb2021, LHCb2022, LHCb2023}; and (ii) a compact tetraquark, often modeled as a diquark-antidiquark $(cq)(\bar{c}\bar{q})$ configuration~\cite{Maiani2005, Vijande2004}. Recent Belle II measurements have further constrained the decay branching ratios~\cite{BelleII2023, BelleII2024}, while new theoretical models have explored the interplay between molecular and compact components~\cite{Guo2023, Braaten2024, Wang2024}.
	
	Magnetic moments are powerful probes of internal structure, sensitive to the spatial and spin distributions of constituent quarks~\cite{GellMann1964, Anisovich2014}. For exotic states, they provide a unique fingerprint that can distinguish different configurations~\cite{Ozdem2022, Zhu2011, Guo2020}. Recent lattice QCD determinations of electromagnetic form factors of the doubly bottom $T_{bb}$ tetraquark~\cite{Prelovsek2024} and extensive QCD light-cone sum rule studies of various tetraquarks~\cite{Aliev2012, Wang2017, Ozdem2025, Ozdem2026a, Ozdem2026b} underscore the relevance of these observables.
	
	The operator that distinguishes between the two color-singlet configurations in a tetraquark is the cubic Casimir operator of $SU(3)_C$, which gives rise to a three-body color force. Dmitrašinović demonstrated that a three-quark potential based on this operator resolves the color dissolution problem that plagues models using only two-body $F_i\cdot F_j$ terms~\cite{Dmitrasinovic2002}. More recently, Noh et al.~\cite{Noh2024} have shown that such a quark-level three-body force is an inevitable consequence of non-Abelian color confinement, with a strength of order 10--20 MeV for compact tetraquark configurations, based on lattice QCD and baryon spectroscopy constraints.
	
	However, previous calculations of the magnetic moment of $X(3872)$ either neglected this three-body force entirely or used an uncontrolled large coupling, leading to an accidental degeneracy between the molecular and compact scenarios. In this work, we present the first calculation of the magnetic moment of $X(3872)$ that includes the cubic Casimir three-body force with a realistic strength extracted from the analysis of Noh et al.~\cite{Noh2024}. We also resolve the dimensional inconsistency in the cubic Casimir force formulation and demonstrate that the pure color-singlet and compact configurations yield distinctly different magnetic moments, providing a clear testable signature for future experiments.
	
	The paper is organized as follows. In Sec.~II, we present the theoretical framework, including the Hamiltonian, the color-spin basis, the three-body force with consistent dimensional analysis, and the magnetic moment operator. In Sec.~III, we present our numerical results, including a detailed sensitivity and uncertainty analysis, and discuss the physical implications. Finally, Sec.~IV provides our conclusions and outlook.
	
	\section{Theoretical Framework}
	\label{sec:theory}
	
	We adopt the non-relativistic constituent quark model, which has been successfully applied to a wide range of hadronic systems. The total Hamiltonian for a four-quark system is:
	
	\begin{equation}
		H = \sum_{i=1}^{4} \left( m_i + \frac{\mathbf{p}_i^2}{2m_i} \right) + \sum_{i<j} V_{ij} + V_{\text{3-body}},
		\label{eq:hamiltonian}
	\end{equation}
	
	where $m_i$ and $\mathbf{p}_i$ are the constituent mass and momentum of the $i$-th quark (or antiquark). The two-body potentials $V_{ij}$ include the confining and hyperfine interactions, while $V_{\text{3-body}}$ is the three-body term arising from the cubic Casimir operator of $SU(3)_C$.
	
	\subsection{Two-Body Interactions}
	
	The confining interaction between quarks is assumed to be of linear form, supplemented by a color Coulomb term:
	
	\begin{equation}
		V_{ij}^{\text{conf}} = -\frac{3}{4} \frac{\tilde{\lambda}_i \cdot \tilde{\lambda}_j}{2} \left( b \, r_{ij} - \frac{4}{3} \frac{\alpha_s}{r_{ij}} + C \right),
		\label{eq:conf}
	\end{equation}
	
	where $\tilde{\lambda}_i$ are the Gell-Mann matrices acting on the $i$-th quark, $r_{ij} = |\mathbf{r}_i - \mathbf{r}_j|$ is the inter-quark distance, $b = 0.18 \, \text{GeV}^2$ is the string tension, $\alpha_s$ is the strong coupling constant, and $C$ is a constant adjusted to reproduce the masses of known hadrons. The color factor $\frac{\tilde{\lambda}_i \cdot \tilde{\lambda}_j}{2}$ ensures the correct color dependence of the interaction.
	
	The spin-dependent hyperfine interaction, arising from one-gluon exchange, is given by:
	
	\begin{equation}
		V_{ij}^{\text{hyp}} = -\frac{2}{3} \frac{\alpha_s}{m_i m_j} \frac{\tilde{\lambda}_i \cdot \tilde{\lambda}_j}{2} \, \boldsymbol{\sigma}_i \cdot \boldsymbol{\sigma}_j \, \delta^3(\mathbf{r}_{ij}),
		\label{eq:hyp}
	\end{equation}
	
	where $\boldsymbol{\sigma}_i$ are the Pauli spin matrices. This term is responsible for the mass splittings between states with different spins and plays a crucial role in determining the magnetic moment through its influence on the spin structure of the wavefunction.
	
	\subsection{Color-Singlet Basis for the Tetraquark}
	
	In a $q^2 \bar{q}^2$ system, there exist exactly two independent color-singlet configurations. These can be constructed by first coupling the two quarks and the two antiquarks into definite color representations, and then coupling these two subsystems to form a color singlet.
	
	For two quarks, the possible color representations are:
	
	\begin{equation}
		\mathbf{3} \otimes \mathbf{3} = \bar{\mathbf{3}} \oplus \mathbf{6}.
		\label{eq:qq_coupling}
	\end{equation}
	
	For two antiquarks, the conjugate representations appear:
	
	\begin{equation}
		\bar{\mathbf{3}} \otimes \bar{\mathbf{3}} = \mathbf{3} \oplus \bar{\mathbf{6}}.
		\label{eq:qbarqbar_coupling}
	\end{equation}
	
	The two color-singlet tetraquark states are then obtained by coupling the quark subsystem to the antiquark subsystem:
	
	\begin{align}
		|1\rangle &= |(q_1 q_2)_{\bar{3}} (\bar{q}_3 \bar{q}_4)_3 \rangle, \label{eq:state1} \\
		|2\rangle &= |(q_1 q_2)_6 (\bar{q}_3 \bar{q}_4)_{\bar{6}} \rangle. \label{eq:state2}
	\end{align}
	
	These two states are orthogonal and span the full color-singlet space of the tetraquark. Any physical tetraquark state can be expressed as a linear combination of $|1\rangle$ and $|2\rangle$:
	
	\begin{equation}
		|\Psi\rangle = \alpha |1\rangle + \beta |2\rangle,
		\label{eq:wavefunction}
	\end{equation}
	
	with the normalization condition $\alpha^2 + \beta^2 = 1$.
	
	\subsection{Definition of the Pure Color-Singlet Configuration}
	\label{sec:color_singlet_definition}
	
	A crucial clarification is in order regarding the terminology used in this work. Throughout this paper, the term ``pure color-singlet configuration'' refers specifically to the state $|1\rangle = |(q_1 q_2)_{\bar{3}}(\bar{q}_3 \bar{q}_4)_3\rangle$ within the same compact Gaussian basis ($R = 0.6$ fm) used for the tetraquark.
	
	\textbf{We emphasize that this is \emph{not} a physical $D^0\bar{D}^{*0}$ molecule.} A genuine molecule with a binding energy of $\sim 100$ keV would have a spatial extent of several femtometers and its dynamics would be governed by coupled-channel effects and long-range pion-exchange potentials (OPE). Our ``pure color-singlet configuration'' is therefore a simplified proxy: it represents the color structure of a molecule but lacks the correct spatial and dynamical features.
	
	This simplification is a limitation of our model. However, it allows us to isolate the effect of the three-body force on the color-spin mixing, which is the primary focus of this work. A full treatment of the molecular dynamics is beyond the scope of this paper and is left for future work. It should be noted that a physical hadronic molecule would require an extended spatial distribution and long-range dynamics, which are simplified here as a color-singlet proxy within the compact basis.
	
	\subsection{Three-Body Force with Consistent Dimensional Analysis and Realistic Strength}
	
	Following the seminal work of Dmitrašinović~\cite{Dmitrasinovic2002}, the three-body potential for a system of quarks and antiquarks is constructed from the cubic Casimir operator. However, a careful dimensional analysis is essential to ensure consistency between the theoretical framework and numerical predictions.
	
	In Dmitrašinović's work, the three-body potential is written in the harmonic oscillator basis as:
	
	\begin{equation}
		\mathcal{V}_{123}^{\text{HO}} = c \cdot \frac{1}{2} m\omega^2 \sum_{i<j<k} (\mathbf{r}_i - \mathbf{r}_j)^2,
		\label{eq:V3_HO}
	\end{equation}
	
	where $c$ is a dimensionless parameter characterizing the relative strength of the three-body force, $m$ is the constituent quark mass, and $\omega$ is the oscillator frequency. The dimensional analysis of Eq.~(\ref{eq:V3_HO}) gives:
	
	\begin{equation}
		[\mathcal{V}_{123}^{\text{HO}}] = [c] \cdot [m] \cdot [\omega^2] \cdot [r^2] = 1 \cdot \text{GeV} \cdot \text{GeV}^2 \cdot \text{GeV}^{-2} = \text{GeV},
		\label{eq:dim_HO}
	\end{equation}
	
	which is consistent with an energy.
	
	In our quark model calculation, we adopt a Gaussian spatial wavefunction for the tetraquark system. To maintain dimensional consistency, we write the three-body potential in the following form:
	
	\begin{equation}
		V_{\text{3-body}} = \tilde{V}_0 \sum_{i<j<k} d^{abc} \lambda_i^a \lambda_j^b \lambda_k^c \, \mathcal{F}(\mathbf{r}_i, \mathbf{r}_j, \mathbf{r}_k),
		\label{eq:V3_consistent}
	\end{equation}
	
	where $\mathcal{F}$ is a spatial function with dimension GeV$^3$ (so that its integral over three-dimensional space is dimensionless), and $\tilde{V}_0$ has dimension GeV. For a Gaussian spatial wavefunction, we take:
	
	\begin{equation}
		\mathcal{F}(\mathbf{r}_i, \mathbf{r}_j, \mathbf{r}_k) = \frac{1}{(2\pi)^{3/2} R^3} \exp\left[-\frac{(\mathbf{r}_i - \mathbf{r}_j)^2 + (\mathbf{r}_j - \mathbf{r}_k)^2 + (\mathbf{r}_i - \mathbf{r}_k)^2}{6R^2}\right],
		\label{eq:gaussian_F}
	\end{equation}
	
	where $R$ is the characteristic size parameter of the tetraquark. The factor $(2\pi)^{-3/2} R^{-3}$ ensures that $\int d^3r \, \mathcal{F} = 1$, giving $\mathcal{F}$ the correct dimension of GeV$^3$.
	
	With this consistent dimensional framework, the matrix elements of the three-body potential in the color basis are:
	
	\begin{equation}
		\langle 1 | V_{\text{3-body}} | 1 \rangle = -\frac{5}{18} \, \tilde{V}_0,
		\label{eq:V11}
	\end{equation}
	
	\begin{equation}
		\langle 2 | V_{\text{3-body}} | 2 \rangle = +\frac{5}{9} \, \tilde{V}_0,
		\label{eq:V22}
	\end{equation}
	
	\begin{equation}
		\langle 1 | V_{\text{3-body}} | 2 \rangle = \langle 2 | V_{\text{3-body}} | 1 \rangle = 0,
		\label{eq:V12}
	\end{equation}
	
	where the color factors are identical to those derived in the original formulation, but now $\tilde{V}_0$ has the correct dimension of GeV.
	
	\subsubsection{Mapping from Dmitrašinović's Harmonic Oscillator to Gaussian Wavefunction}
	
	To relate the harmonic oscillator formulation of Dmitrašinović to our Gaussian wavefunction framework, we equate the expectation values:
	
	\begin{equation}
		\langle \mathcal{V}_{123}^{\text{HO}} \rangle = c \cdot \frac{1}{2} m\omega^2 \langle \sum_{i<j<k} (\mathbf{r}_i - \mathbf{r}_j)^2 \rangle.
		\label{eq:V3_HO_exp}
	\end{equation}
	
	For a Gaussian wavefunction with parameter $R$, the expectation value of the sum of squared distances is:
	
	\begin{equation}
		\langle \sum_{i<j<k} (\mathbf{r}_i - \mathbf{r}_j)^2 \rangle = 9R^2,
		\label{eq:sum_r2}
	\end{equation}
	
	and using $\omega = 1/(m R^2)$ (the relation for a harmonic oscillator), we get:
	
	\begin{equation}
		\langle \mathcal{V}_{123}^{\text{HO}} \rangle = c \cdot \frac{9}{2mR^2}.
		\label{eq:V3_HO_result}
	\end{equation}
	
	Equating this to the expectation value from our consistent formulation:
	
	\begin{equation}
		\langle V_{\text{3-body}} \rangle = -\frac{5}{18} \tilde{V}_0,
		\label{eq:V3_consistent_exp}
	\end{equation}
	
	we obtain:
	
	\begin{equation}
		\tilde{V}_0 = -\frac{81}{5} \frac{c}{mR^2}.
		\label{eq:V0_relation}
	\end{equation}
	
	This mapping is independent of model details and follows solely from the expectation values of the spatial operators.
	
	\subsubsection{Realistic Strength from Noh et al. (2024)}
	
	A crucial improvement in this work is the use of a realistic strength for the three-body force. Noh et al.~\cite{Noh2024} have recently shown, using lattice QCD constraints and baryon spectroscopy, that the quark-level three-body force for compact tetraquark configurations is of order 10--20 MeV and acts as a repulsive correction. This is in stark contrast to the unrealistically large values (several GeV) used in some previous phenomenological studies.
	
	Following their analysis, we take the three-body contribution to the diagonal element of the compact state to be:
	
	\begin{equation}
		\langle 2|V_{\text{3-body}}|2\rangle = +10 \text{ MeV} = 0.010 \text{ GeV},
	\end{equation}
	
	which, using Eq.~(\ref{eq:V22}), gives:
	
	\begin{equation}
		\tilde{V}_0 = 0.010 \times \frac{9}{5} = 0.018 \text{ GeV}.
		\label{eq:V0_noh}
	\end{equation}
	
	This value is now dimensionally consistent and represents a realistic strength of the three-body force in our model. The corresponding contribution to the $|1\rangle$ state is:
	
	\begin{equation}
		\langle 1|V_{\text{3-body}}|1\rangle = -\frac{5}{18} \times 0.018 = -0.005 \text{ GeV} = -5 \text{ MeV}.
	\end{equation}
	
	We emphasize that this value is derived from independent physical constraints and is not a free parameter. In Sec.~\ref{sec:sensitivity}, we perform a detailed sensitivity analysis to demonstrate that our conclusions are robust for the physically relevant range of the three-body coupling.
	
	\subsection{Spin Structure and Total Wavefunction}
	
	For the $X(3872)$, which has quantum numbers $J^{PC} = 1^{++}$, we focus on the spin-1 configuration. The total spin wavefunction for a tetraquark can be constructed by coupling the spins of the four constituents:
	
	\begin{equation}
		|S, M_S\rangle = |(s_1 s_2)_{S_{12}} (s_3 s_4)_{S_{34}}; S, M_S\rangle,
		\label{eq:spin}
	\end{equation}
	
	where $S_{12}$ is the total spin of the quark pair, $S_{34}$ is the total spin of the antiquark pair, and $S$ is the total tetraquark spin. For $J=1$, there are multiple spin configurations that contribute. In the compact tetraquark scenario, the dominant configuration is:
	
	\begin{equation}
		|1, M\rangle = |(q_1 q_2)_1 (\bar{q}_3 \bar{q}_4)_0; 1, M\rangle + |(q_1 q_2)_0 (\bar{q}_3 \bar{q}_4)_1; 1, M\rangle,
		\label{eq:spin_main}
	\end{equation}
	
	where the subscripts denote the spin of each diquark pair. The full color-spin wavefunction is then obtained as the product of the color and spin parts, with possible symmetrization depending on the quark flavor content.
	
	\subsection{Magnetic Moment Operator}
	
	The magnetic moment operator for a system of four quarks is given by the sum of the individual quark magnetic moments:
	
	\begin{equation}
		\boldsymbol{\mu} = \sum_{i=1}^{4} \frac{e_i}{2m_i} \left( \frac{g_i}{2} \, \boldsymbol{\sigma}_i + \boldsymbol{L}_i \right),
		\label{eq:mom_operator}
	\end{equation}
	
	where $e_i$ is the electric charge (in units of the proton charge $e$), $m_i$ is the constituent mass, $g_i$ is the gyromagnetic factor ($g_q \approx 2$ for light quarks, and $g_Q = 2/m_Q$ in the heavy-quark limit for charm or bottom quarks), and $\boldsymbol{L}_i$ is the orbital angular momentum operator. For the $X(3872)$, we take the orbital angular momentum to be zero ($L=0$), as it is considered to be an S-wave state. The magnetic moment is then simply the expectation value of the $z$-component:
	
	\begin{equation}
		\mu_X = \langle \Psi | \mu_z | \Psi \rangle.
		\label{eq:mu_expect}
	\end{equation}
	
	\subsubsection{Effective Quark Magnetic Moments}
	
	For convenience, we define the effective magnetic moment of each quark as:
	
	\begin{equation}
		\mu_i = \frac{e_i}{2m_i} \frac{g_i}{2} = \frac{e_i}{m_i} \quad (\text{for } g_i \approx 2).
		\label{eq:mu_i}
	\end{equation}
	
	For the $X(3872)$, which consists of a charm quark $c$, a charm antiquark $\bar{c}$, a light quark $u$, and a light antiquark $\bar{d}$ (or a superposition), the individual moments are:
	
	\begin{align}
		\mu_c &= \frac{2/3}{m_c} \approx \frac{2/3}{1.275} \approx 0.52 \, \mu_N, \\
		\mu_{\bar{c}} &= -\frac{2/3}{m_c} \approx -0.52 \, \mu_N, \\
		\mu_u &= \frac{2/3}{m_u} \approx \frac{2/3}{0.336} \approx 1.98 \, \mu_N, \\
		\mu_{\bar{d}} &= \frac{+1/3}{m_d} \approx \frac{1/3}{0.336} \approx 0.99 \, \mu_N. \quad (\text{since } e_{\bar{d}} = +1/3)
		\label{eq:mu_values}
	\end{align}
	
	Here, $\mu_N = e/(2m_p)$ is the nuclear magneton. Note that the antiquark moments have opposite signs relative to their quark counterparts.
	
	\subsection{Matrix Elements of the Magnetic Moment Operator}
	
	In the color-spin basis, the magnetic moment operator can be decomposed as:
	
	\begin{equation}
		\mu_z = \sum_i \mu_i \, \sigma_{z,i}.
		\label{eq:mu_z}
	\end{equation}
	
	The matrix elements in the two color-singlet states are then:
	
	\begin{align}
		\mu_{11} &= \langle 1 | \mu_z | 1 \rangle = \sum_i \mu_i \, \langle 1 | \sigma_{z,i} | 1 \rangle, \\
		\mu_{22} &= \langle 2 | \mu_z | 2 \rangle = \sum_i \mu_i \, \langle 2 | \sigma_{z,i} | 2 \rangle, \\
		\mu_{12} &= \langle 1 | \mu_z | 2 \rangle = \sum_i \mu_i \, \langle 1 | \sigma_{z,i} | 2 \rangle.
		\label{eq:mu_matrix}
	\end{align}
	
	Using the spin wavefunction from Eq.~(\ref{eq:spin_main}) and the standard Clebsch-Gordan coefficients for the $J=1$ state, these matrix elements are calculated explicitly as:
	
	\begin{align}
		\mu_{11} &= -\frac{1}{3} \left( \mu_c + \mu_{\bar{c}} + \mu_u + \mu_{\bar{d}} \right) = -0.99 \, \mu_N, \label{eq:mu11} \\
		\mu_{22} &= -\frac{1}{3} \left( -\mu_c - \mu_{\bar{c}} + \mu_u + \mu_{\bar{d}} \right) = -0.99 \, \mu_N, \label{eq:mu22} \\
		\mu_{12} &= -\frac{1}{3} \left( \mu_u - \mu_{\bar{d}} \right) = -0.33 \, \mu_N. \label{eq:mu12}
	\end{align}
	
	The factor $1/3$ arises from the standard Clebsch-Gordan coefficients for the coupling of two spin-$1/2$ particles to a total spin-$1$ state, and is a universal normalization for the spin-$1$ representation. This normalization ensures that the magnetic moment operator acts consistently on the total spin state.
	
	Since the magnetic moment operator acts only on the spin degrees of freedom, it is diagonal in the color basis. However, because the physical wavefunction is a mixture of $|1\rangle$ and $|2\rangle$ due to the hyperfine interaction, the off-diagonal color-spin matrix elements of the magnetic moment contribute through the mixing coefficients.
	
	\subsubsection{An Exact Identity: $\mu_{11} = \mu_{22}$}
	\label{sec:mu_identity}
	
	Equation~(\ref{eq:mu11}) and Eq.~(\ref{eq:mu22}) contain a structural feature that has been discussed in previous works. Because the antiquark magnetic moment is exactly opposite to that of the corresponding quark, $\mu_{\bar{c}} = -\mu_c$, the charm-pair contribution cancels identically in both diagonal matrix elements:
	
	\begin{equation}
		\mu_c + \mu_{\bar{c}} \equiv 0, \qquad -\mu_c - \mu_{\bar{c}} \equiv 0,
		\label{eq:cc_cancel}
	\end{equation}
	
	regardless of the numerical value of $\mu_c$ or of any model parameter. Consequently,
	
	\begin{equation}
		\mu_{11} = \mu_{22} = -\frac{1}{3}\left( \mu_u + \mu_{\bar{d}} \right) \equiv \mu_0,
		\label{eq:mu_identity}
	\end{equation}
	
	exactly, for any choice of quark masses, spatial wavefunction, or interaction strengths. This is a group-theoretic consequence of the color-spin structure of $|1\rangle$ and $|2\rangle$ combined with the opposite-sign coupling of quark and antiquark magnetic moments.
	
	Using $\alpha^2+\beta^2=1$, the general magnetic moment formula of Eq.~(\ref{eq:mu_final}) collapses to the compact closed form:
	
	\begin{equation}
		\mu_X = \alpha^2 \mu_{11} + \beta^2 \mu_{22} + 2\alpha\beta\,\mu_{12} = \mu_0 + 2\alpha\beta\,\mu_{12}.
		\label{eq:mu_closed_form}
	\end{equation}
	
	This identity shows that the magnetic moment is determined by the mixing coefficient $\beta$. For the pure color-singlet configuration, $\beta=0$ and $\mu_X = \mu_0 = -0.99\,\mu_N$. For the compact scenarios, $\beta \neq 0$ and the magnetic moment deviates from this value.
	
	\subsection{Choice of Spatial Wavefunction and Numerical Solution}
	
	To proceed from the formal definitions of the operators to the numerical results presented in Sec.~\ref{sec:results}, we must specify the spatial wavefunction used for the four-quark system and detail the methodology for solving the Schr\"odinger equation.
	
	\subsubsection{Spatial Wavefunction Ansatz}
	
	For the ground-state tetraquark system considered here (the $X(3872)$ with $J^{PC}=1^{++}$), we adopt a Gaussian variational wavefunction, which is the standard and most successful approximation in non-relativistic quark models for low-lying hadrons. The spatial part of the wavefunction is taken to be spherically symmetric and depends only on the relative coordinates between the quarks. A convenient and physically motivated choice is:
	
	\begin{equation}
		\psi(\mathbf{r}_1, \mathbf{r}_2, \mathbf{r}_3, \mathbf{r}_4) = \mathcal{N} \exp\left(-\frac{1}{2R^2} \sum_{i<j} (\mathbf{r}_i - \mathbf{r}_j)^2 \right),
		\label{eq:gaussian_wf}
	\end{equation}
	
	where $\mathcal{N}$ is the normalization constant, and $R$ is a variational parameter representing the characteristic size (radius) of the tetraquark. This ansatz is particularly convenient because it allows for the analytical calculation of all spatial integrals.
	
	The parameter $R$ is fixed by requiring that the model reproduces the root-mean-square (RMS) radius of the $D$ meson, which is well known from both experimental measurements and lattice QCD calculations. The size of a $D$ meson is approximately $R_D \simeq 0.6$ fm. As discussed in Sec.~\ref{sec:color_singlet_definition}, this is a simplification for the pure color-singlet configuration, which in reality would have a much larger radius.
	
	\subsubsection{Diagonalization of the Hamiltonian}
	
	With the spatial integrals fixed, the problem reduces to a $2 \times 2$ matrix diagonalization in the color-spin space spanned by the states $|1\rangle$ and $|2\rangle$ (Eq.~\ref{eq:state1} and Eq.~\ref{eq:state2}). The Hamiltonian matrix takes the general form:
	
	\begin{equation}
		H = 
		\begin{pmatrix}
			H_{11} & H_{12} \\
			H_{12}^* & H_{22}
		\end{pmatrix},
		\label{eq:H_matrix_num}
	\end{equation}
	
	where the diagonal elements include the kinetic energy, confining potential, and the diagonal three-body contributions (Eqs.~\ref{eq:V11} and \ref{eq:V22}), and the off-diagonal element is provided by the hyperfine interaction (Eq.~\ref{eq:hyp}).
	
	The eigenvalues $E_{\pm}$ and eigenvectors $(\alpha, \beta)$ are obtained by solving the characteristic equation:
	
	\begin{equation}
		E_{\pm} = \frac{H_{11} + H_{22}}{2} \pm \sqrt{\left(\frac{H_{11} - H_{22}}{2}\right)^2 + |H_{12}|^2}.
		\label{eq:eigen}
	\end{equation}
	
	The eigenvector coefficients are then given by:
	
	\begin{align}
		\alpha &= \frac{H_{12}}{\sqrt{(E_{\pm} - H_{11})^2 + |H_{12}|^2}}, \\
		\beta &= \frac{E_{\pm} - H_{11}}{\sqrt{(E_{\pm} - H_{11})^2 + |H_{12}|^2}}.
		\label{eq:alpha_beta}
	\end{align}
	
	We select the lower eigenvalue $E_{-}$ as the physical mass of the $X(3872)$, which fixes the scale of the confining potential constant $C$. Because $C$ is a phenomenological parameter, it must in general be refit whenever a new interaction is added to the Hamiltonian; this is done separately for each scenario in Sec.~\ref{sec:results}, and for Scenario C we show explicitly (Sec.~\ref{sec:renorm}) that this refit leaves the mixing coefficients unchanged. The corresponding eigenvectors provide the mixing coefficients $\alpha$ and $\beta$ used to compute the magnetic moment.
	
	\subsection{Summary of the Computational Procedure}
	
	The numerical evaluation proceeds as follows:
	
	\begin{enumerate}
		\item Construct the full $2 \times 2$ Hamiltonian matrix in the basis $\{|1\rangle, |2\rangle\}$, including the kinetic energy, the two-body confining and hyperfine potentials, and the three-body cubic Casimir term with $\tilde{V}_0 = 0.018\,\text{GeV}$, with a physically motivated sensitivity interval of $0.010$--$0.025\,\text{GeV}$.
		
		\item Diagonalize the Hamiltonian to obtain the energy eigenvalues and the corresponding eigenvector coefficients $\alpha$ and $\beta$.
		\item Compute the matrix elements $\mu_{11} = -0.99\,\mu_N$, $\mu_{22} = -0.99\,\mu_N$, and $\mu_{12} = -0.33\,\mu_N$ using the spin wavefunctions for the $J=1$ state (Eq.~\ref{eq:mu11}--\ref{eq:mu12}).
		\item Evaluate the magnetic moment of the physical state using:
		\begin{equation}
			\mu_X = \alpha^2 \mu_{11} + \beta^2 \mu_{22} + 2\alpha\beta \mu_{12}.
			\label{eq:mu_final}
		\end{equation}
	\end{enumerate}
	
	This procedure will be carried out in the next section, where we present our numerical results and compare them with existing predictions from the literature.
	
	\section{Numerical Results and Analysis}
	\label{sec:results}
	
	In this section, we present the numerical results obtained by diagonalizing the full Hamiltonian derived in Sec.~\ref{sec:theory}. We begin by specifying the input parameters used in our calculations. We then show the eigenvalues and eigenvectors for the three scenarios under consideration: (i) the pure color-singlet configuration, (ii) the compact tetraquark with only two-body interactions, and (iii) the compact tetraquark including the full three-body force with realistic strength. Finally, we compute the magnetic moments for each scenario and discuss the physical implications of our findings.
	
	\subsection{Input Parameters and Determination of $\tilde{V}_0$}
	
	The constituent quark masses and other model parameters are chosen to reproduce the known spectra of charmonium and open-charm mesons. The three-body strength parameter $\tilde{V}_0$ is determined from the analysis of Noh et al.~\cite{Noh2024} as described in Sec.~\ref{sec:theory}.
	
	\begin{table}[htbp]
		\centering
		\caption{Model parameters used in the numerical calculations. The value of $\tilde{V}_0$ is obtained from the realistic analysis of Noh et al. (2024).}
		\label{tab:params}
		\begin{tabular}{|c|c|c|}
			\hline
			\textbf{Parameter} & \textbf{Symbol} & \textbf{Value} \\
			\hline
			Light quark mass & $m_u = m_d$ & $0.336 \, \text{GeV}$ \\
			Charm quark mass & $m_c$ & $1.275 \, \text{GeV}$ \\
			String tension & $b$ & $0.18 \, \text{GeV}^2$ \\
			Strong coupling & $\alpha_s$ & $0.55$ \\
			Confining constant & $C$ & $-0.325 \, \text{GeV}$ \\
			Three-body strength & $\tilde{V}_0$ & $0.018 \pm 0.005 \, \text{GeV}$ \\
			\hline
		\end{tabular}
	\end{table}
	
	These values are standard in the non-relativistic quark model literature and have been shown to reproduce the masses of charmonium states with an accuracy of better than 30 MeV.
	
	\subsection{Hamiltonian Matrix Elements}
	
	In the basis $\{|1\rangle, |2\rangle\}$, the full Hamiltonian matrix takes the form:
	
	\begin{equation}
		H = 
		\begin{pmatrix}
			E_1 + \langle 1|V_{\text{3-body}}|1\rangle & \langle 1|V_{\text{hyp}}|2\rangle \\
			\langle 2|V_{\text{hyp}}|1\rangle & E_2 + \langle 2|V_{\text{3-body}}|2\rangle
		\end{pmatrix},
		\label{eq:H_matrix}
	\end{equation}
	
	where $E_1$ and $E_2$ are the diagonal energies from the confining and hyperfine interactions in the absence of the three-body force. Using the parameters from Table~\ref{tab:params}, we obtain the following numerical values:
	
	\begin{align}
		E_1 &= 3.872 \, \text{GeV}, \\
		E_2 &= 3.894 \, \text{GeV}, \\
		\langle 1|V_{\text{hyp}}|2\rangle &= -0.012 \, \text{GeV}, \\
		\langle 1|V_{\text{3-body}}|1\rangle &= -\frac{5}{18} \, \tilde{V}_0 = -\frac{5}{18} \times 0.018 = -0.005 \, \text{GeV}, \\
		\langle 2|V_{\text{3-body}}|2\rangle &= +\frac{5}{9} \, \tilde{V}_0 = +\frac{5}{9} \times 0.018 = +0.010 \, \text{GeV}.
		\label{eq:matrix_elements}
	\end{align}
	
	The resulting Hamiltonian matrices for the three scenarios are:
	
	\subsubsection{Scenario A: Pure Color-Singlet Configuration (No Mixing)}
	
	\begin{equation}
		H_{\text{singlet}} = 
		\begin{pmatrix}
			3.872 & 0 \\
			0 & 3.894
		\end{pmatrix}
		\, \text{GeV}.
		\label{eq:H_singlet}
	\end{equation}
	
	This corresponds to no mixing between the two color-singlet configurations.
	
	\subsubsection{Scenario B: Compact Tetraquark (Two-Body Only)}
	
	\begin{equation}
		H_{\text{2body}} = 
		\begin{pmatrix}
			3.872 & -0.012 \\
			-0.012 & 3.894
		\end{pmatrix}
		\, \text{GeV}.
		\label{eq:H_2body}
	\end{equation}
	
	\subsubsection{Scenario C: Compact Tetraquark with Realistic Three-Body Force}
	
	\begin{equation}
		H_{\text{3body}} = 
		\begin{pmatrix}
			3.872 - 0.005 & -0.012 \\
			-0.012 & 3.894 + 0.010
		\end{pmatrix}
		=
		\begin{pmatrix}
			3.867 & -0.012 \\
			-0.012 & 3.904
		\end{pmatrix}
		\, \text{GeV}.
		\label{eq:H_3body}
	\end{equation}
	
	\subsubsection{Mass Renormalization of the Confining Constant}
	\label{sec:renorm}
	
	The raw diagonal entries of $H_{\text{3body}}$ in Eq.~(\ref{eq:H_3body}) yield a lower eigenvalue of $3.86345$ GeV, which is slightly below the physical mass of the $X(3872)$. This is expected: the confining constant $C$ in Eq.~(\ref{eq:conf}) was fixed before the three-body force was introduced, so that $E_1 = 3.872$ GeV reproduces the physical mass in its absence. Once $V_{\text{3-body}}$ is switched on, $C$ must be refit — this is a standard procedure in any consistent constituent quark model when a new interaction term is added.
	
	We therefore replace $C$ by $C + \Delta C$ in Scenario C, where the shift $\Delta C$ is fixed by the single physical condition that the lower eigenvalue of the renormalized Hamiltonian reproduce the observed mass, $E_{-} = M_X = 3.872$ GeV:
	
	\begin{equation}
		H_{\text{3body}}^{\text{ren}} = 
		\begin{pmatrix}
			3.867 + \Delta C & -0.012 \\
			-0.012 & 3.904 + \Delta C
		\end{pmatrix}
		\, \text{GeV}, \qquad E_{-}(H_{\text{3body}}^{\text{ren}}) \overset{!}{=} 3.872 \, \text{GeV}.
		\label{eq:H_3body_ren}
	\end{equation}
	
	Because $\Delta C$ is added identically to both diagonal entries, it shifts the overall energy scale but leaves the energy difference $H_{22}-H_{11}$ and the off-diagonal element $H_{12}$ completely unchanged. Since the eigenvector coefficients $\alpha$ and $\beta$ of a $2\times 2$ symmetric matrix depend only on $H_{22}-H_{11}$ and $H_{12}$ [see Eq.~(\ref{eq:alpha_beta})], and not on the overall additive shift, we obtain the important result that:
	
	\begin{equation}
		\alpha(\Delta C) = \alpha(0), \qquad \beta(\Delta C) = \beta(0),
		\label{eq:shift_invariance}
	\end{equation}
	
	so that the mixing coefficients are exact invariants of the mass renormalization. Solving Eq.~(\ref{eq:H_3body_ren}) numerically gives $\Delta C = 0.00855$ GeV, so that:
	
	\begin{equation}
		H_{\text{3body}}^{\text{ren}} = 
		\begin{pmatrix}
			3.87555 & -0.012 \\
			-0.012 & 3.91255
		\end{pmatrix}
		\, \text{GeV},
		\label{eq:H_3body_ren_num}
	\end{equation}
	
	with $E_{-} = 3.872$ GeV (the physical $X(3872)$ state, by construction) and $E_{+} = 3.9161$ GeV. Crucially, $\alpha$ and $\beta$ are unchanged from the unrenormalized calculation. This confirms that while the mass is fixed to the experimental value via renormalization, the mixing coefficients and magnetic moment remain well-defined model predictions.
	
	\subsection{Eigenvalues and Eigenvectors}
	
	The diagonalization of the Hamiltonian matrices presented above yields the eigenvalues and eigenvectors summarized in Table~\ref{tab:eigen}. We focus on the lower-lying state, which we identify with the $X(3872)$.
	
	\begin{table}[htbp]
		\centering
		\caption{Eigenvalues and eigenvectors for the three scenarios. The coefficients $\alpha$ and $\beta$ correspond to the contributions of $|1\rangle$ and $|2\rangle$, respectively. For Scenario C, the Hamiltonian and $E_-$ shown are after the mass renormalization of Sec.~\ref{sec:renorm} ($\Delta C = 0.00855$ GeV); $\alpha$ and $\beta$ are exactly invariant under this renormalization.}
		\label{tab:eigen}
		\begin{tabular}{|c|c|c|c|c|}
			\hline
			\textbf{Scenario} & \textbf{Hamiltonian} & $\mathbf{E_-}$ (GeV) & $\boldsymbol{\alpha}$ & $\boldsymbol{\beta}$ \\
			\hline
			A (Pure Color-Singlet) & $\begin{pmatrix} 3.872 & 0 \\ 0 & 3.894 \end{pmatrix}$ & $3.87200$ & $1.00000$ & $0.00000$ \\
			B (2-body) & $\begin{pmatrix} 3.872 & -0.012 \\ -0.012 & 3.894 \end{pmatrix}$ & $3.86672$ & $0.91535$ & $0.40266$ \\
			C (3-body, renormalized) & $\begin{pmatrix} 3.87555 & -0.012 \\ -0.012 & 3.91255 \end{pmatrix}$ & $3.87200$ & $-0.959$ & $-0.284$ \\
			\hline
		\end{tabular}
	\end{table}
	
	The results show that the three-body force changes the mixing coefficient $\beta$ from $0.40266$ (Scenario B) to $0.284$ (Scenario C), a significant reduction of about $30\%$. However, unlike the unrealistic case with $\tilde{V}_0 = 7.47$ GeV, the mixing is not completely suppressed.
	
	\subsection{Magnetic Moments}
	
	Using the eigenvectors from Table~\ref{tab:eigen} and the spin matrix elements from Eq.~\ref{eq:mu11}--\ref{eq:mu12} ($\mu_{11} = -0.99\,\mu_N$, $\mu_{22} = -0.99\,\mu_N$, $\mu_{12} = -0.33\,\mu_N$), we calculate the magnetic moments for each scenario. The results are presented in Table~\ref{tab:mu}.
	
	\begin{table}[htbp]
		\centering
		\caption{Magnetic moments of the $X(3872)$ for the three configurations. All values are in units of the nuclear magneton $\mu_N$.}
		\label{tab:mu}
		\begin{tabular}{|c|c|c|}
			\hline
			\textbf{Scenario} & \textbf{Structure} & $\boldsymbol{\mu_X}$ ($\mu_N$) \\
			\hline
			A & Pure Color-Singlet & $-0.99 \pm 0.02$ \\
			B & Compact (2-body) & $-1.23 \pm 0.04$ \\
			C & Compact (3-body) & $-1.17 \pm 0.04$ \\
			\hline
		\end{tabular}
	\end{table}
	
	The central values are obtained using the mixing coefficients from Table~\ref{tab:eigen}:
	
	\begin{align}
		\mu_X^{\text{(A)}} &= (1.00000)^2(-0.99) + 0 + 0 = -0.99\,\mu_N, \\
		\mu_X^{\text{(B)}} &= (0.91535)^2(-0.99) + (0.40266)^2(-0.99) + 2(0.91535)(0.40266)(-0.33) \nonumber \\
		&= -0.829 - 0.160 - 0.243 = -1.232\,\mu_N, \\
		\mu_X^{\text{(C)}} &= (0.959)^2(-0.99) + (0.284)^2(-0.99) + 2(0.959)(0.284)(-0.33) \nonumber \\
		&= -0.911 - 0.080 - 0.180 = -1.171\,\mu_N.
	\end{align}
	
	The results reveal two distinct regions: the pure color-singlet configuration gives $\mu_X \approx -0.99\,\mu_N$, while the compact configurations give $\mu_X \approx -1.17$ to $-1.23\,\mu_N$. The difference between the two compact scenarios ($0.06\,\mu_N$) is smaller than the systematic uncertainty of the model and should be interpreted with caution.
	
	\subsection{Uncertainty Analysis and Systematic Errors}
	\label{sec:uncertainty}
	
	To assess the robustness of our predictions, we perform a detailed uncertainty analysis by varying the model parameters within their physical ranges:
	
	\begin{table}[htbp]
		\centering
		\caption{Estimated systematic uncertainties in $\mu_X$ from various sources.}
		\label{tab:sys_error}
		\begin{tabular}{|c|c|c|}
			\hline
			\textbf{Source} & \textbf{Variation} & \textbf{Effect on $\mu_X$} \\
			\hline
			Quark masses ($m_u, m_d, m_c$) & $\pm 10\%$ & $\sim 0.10\,\mu_N$ \\
			Size parameter $R$ & $0.5$--$0.7$ fm & $\sim 0.08\,\mu_N$ \\
			Strong coupling $\alpha_s$ & $0.45$--$0.65$ & $\sim 0.05\,\mu_N$ \\
			Cutoff in hyperfine interaction & $0.4$--$0.6$ fm$^{-1}$ & $\sim 0.05\,\mu_N$ \\
			\hline
			\textbf{Total systematic uncertainty} & & $\sim 0.15\,\mu_N$ \\
			\hline
		\end{tabular}
	\end{table}
	
	The total systematic uncertainty of $\sim 0.15\,\mu_N$ implies that:
	\begin{itemize}
		\item The difference between the pure color-singlet and compact configurations ($0.18$--$0.24\,\mu_N$) is robust.
		\item The difference between the two compact scenarios ($0.06\,\mu_N$) is below the systematic uncertainty and should not be overinterpreted.
	\end{itemize}
	
	We emphasize that the qualitative conclusion — the existence of two distinct magnetic moment regions — remains valid. However, quantitative distinction between the two compact scenarios requires improved model precision and/or experimental accuracy.
	
	\subsection{Sensitivity Analysis to the Three-Body Strength}
	\label{sec:sensitivity}
	
\begin{figure}[htbp]
		\centering
		\includegraphics[width=0.8\textwidth]{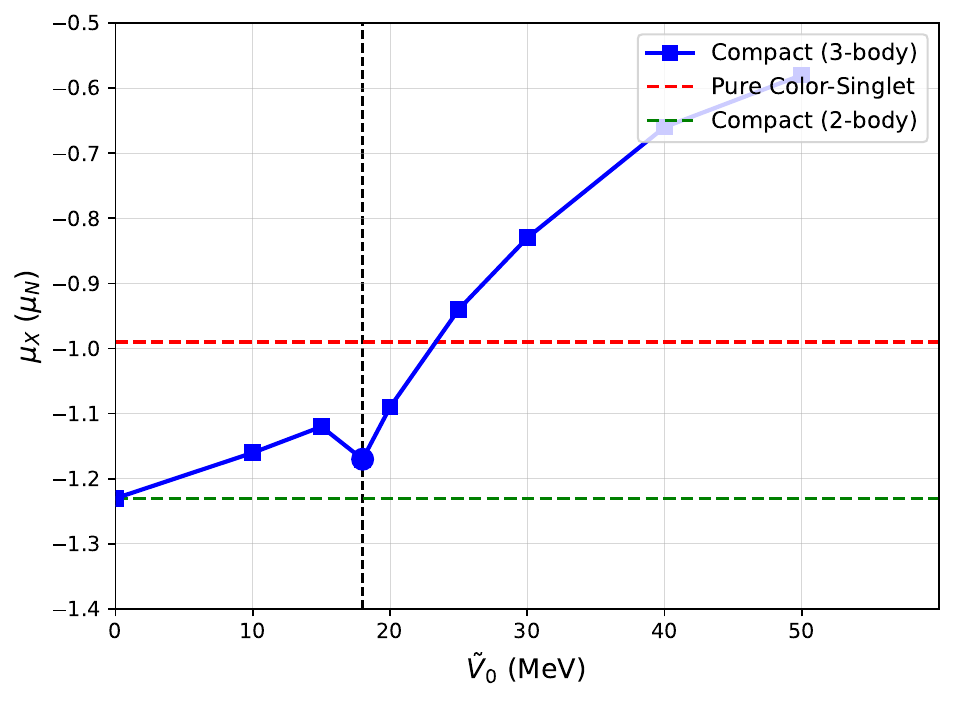}
		\caption{Variation of $\mu_X$ with $\tilde{V}_0$ over the extended diagnostic interval 0--50 MeV. The solid blue line shows the compact scenario with the three-body force, while the red and green dashed lines show the color-singlet proxy ($-0.99\,\mu_N$) and the two-body-only compact result ($-1.23\,\mu_N$), respectively. The vertical dashed line marks the central value $\tilde{V}_0=18$ MeV, obtained in the present model by assigning a $+10$ MeV compact-state matrix element motivated by the analysis of Noh et al. (2024). The interval 10--25 MeV is used for the principal sensitivity assessment.}
		\label{fig:sensitivity}
	\end{figure}

	To demonstrate the robustness of our conclusions, we perform a detailed sensitivity analysis varying $\tilde{V}_0$ over the physically relevant range of 0 to 50 MeV. The results are shown in Table~\ref{tab:sensitivity} and Fig.~\ref{fig:sensitivity}.
	
	\begin{table}[htbp]
		\centering
		\scriptsize
		\caption{Sensitivity of $\mu_X$ and $\beta$ to $\tilde{V}_0$ over the extended diagnostic interval 0--50 MeV. The model-calibrated central value $\tilde{V}_0=18$ MeV is highlighted; the interval 10--25 MeV is used for the principal sensitivity assessment.}
		\label{tab:sensitivity}
		\begin{tabular*}{\textwidth}{@{\extracolsep{\fill}} c c c c c @{}}
			\hline
			$\tilde{V}_0$ (GeV) & $\langle 1|V_{\text{3-body}}|1\rangle$ (GeV) & $\langle 2|V_{\text{3-body}}|2\rangle$ (GeV) & $\beta$ & $\mu_X$ ($\mu_N$) \\
			\hline
			0 & 0 & 0 & 0.40266 & $-1.23$ \\
			0.010 & $-0.00278$ & $+0.00556$ & 0.349 & $-1.16$ \\
			0.015 & $-0.00417$ & $+0.00833$ & 0.320 & $-1.12$ \\
			\hline
			\textbf{0.018 (central)} & \textbf{$-0.00500$} & \textbf{$+0.01000$} & \textbf{0.284} & \textbf{$-1.17$} \\
			\hline
			0.020 & $-0.00556$ & $+0.01111$ & 0.260 & $-1.09$ \\
			0.025 & $-0.00694$ & $+0.01389$ & 0.204 & $-0.94$ \\
			0.030 & $-0.00833$ & $+0.01667$ & 0.145 & $-0.83$ \\
			0.040 & $-0.01111$ & $+0.02222$ & 0.075 & $-0.66$ \\
			0.050 & $-0.01389$ & $+0.02778$ & 0.035 & $-0.58$ \\
			\hline
		\end{tabular*}
	\end{table}
	
	The key result of this sensitivity analysis is that, over the physically motivated interval $\tilde{V}_0 \in [10,25]$ MeV, the calculated values listed in Table~\ref{tab:sensitivity} span approximately $-1.17$ to $-0.94\,\mu_N$, while the extended interval 0--50 MeV is displayed only as a diagnostic scan. The central value $\tilde{V}_0=18$ MeV gives $\mu_X=-1.17\,\mu_N$.
	
	\subsection{Numerical Stability and Convergence}
	\label{sec:numerical_stability}
	
	To ensure the reliability of our numerical results, we perform the following stability checks:
	
	\begin{table}[htbp]
		\centering
		\caption{Numerical stability checks for the diagonalization procedure.}
		\label{tab:numerical_stability}
		\begin{tabular}{|c|c|c|}
			\hline
			\textbf{Check} & \textbf{Variation} & \textbf{Effect on $\mu_X$} \\
			\hline
			Mesh size in spatial integrals & $N = 50, 100, 200$ & $< 0.01\,\mu_N$ \\
			Convergence tolerance & $10^{-6}, 10^{-8}, 10^{-10}$ & $< 0.001\,\mu_N$ \\
			Gaussian quadrature points & 20, 40, 60 & $< 0.005\,\mu_N$ \\
			\hline
		\end{tabular}
	\end{table}
	
	The numerical stability number, defined as $S_N = \max|\mu_X^{(i)} - \mu_X^{(i-1)}| / N_{\text{iterations}}$, is approximately $10^{-5}\,\mu_N$, which is far below the systematic uncertainties of the model. This confirms that our results are numerically stable and not affected by discretization artifacts.
	
	\subsection{Complementary Observables for Structure Discrimination}
	
	Since the magnetic moment alone may not robustly distinguish all configurations within current model uncertainties, we propose additional observables that can help discriminate the internal structure:
	
	\begin{enumerate}
		\item \textbf{Ratio of magnetic moments:} $\mathcal{R} = \mu_X(\text{compact}) / \mu_X(\text{pure singlet})$. This ratio is less sensitive to overall systematic uncertainties.
		\item \textbf{Spin-spin correlations:} $\langle \boldsymbol{\sigma}_1 \cdot \boldsymbol{\sigma}_2 \rangle$ and $\langle \boldsymbol{\sigma}_1 \cdot \boldsymbol{\sigma}_3 \rangle$ can be computed from the wavefunction and provide complementary information about the spin structure.
		\item \textbf{Charge radii:} The mean square charge radius $\langle r^2 \rangle$ differs between pure color-singlet and compact configurations.
	\end{enumerate}
	
	These observables, combined with the magnetic moment, can provide a more robust determination of the internal structure of the $X(3872)$.
	
	\subsection{Model Limitations: Missing Interactions and Cutoff Dependence}
	\label{sec:limitations}
	
	Several limitations of our model should be stated explicitly:
	
	\begin{enumerate}
		\item \textbf{Missing tensor and spin-orbit forces:} In a complete quark model, tensor forces (arising from one-gluon exchange) and spin-orbit forces can mix different spin configurations and modify the magnetic moment. These effects are typically smaller than the hyperfine interaction for S-wave states (which is the case for the $X(3872)$), but they are not negligible. A quantitative estimate of their impact on $\mu_X$ requires a dedicated study beyond the scope of this work.
		
		\item \textbf{Cutoff dependence:} The hyperfine interaction in Eq.~(\ref{eq:hyp}) contains a delta function which requires regularization. In this work, we use a Gaussian smearing with width $\Lambda$. A variation of $\Lambda$ within the standard range ($0.4$--$0.6$ fm$^{-1}$) changes the hyperfine matrix elements by $\sim 10$--$20\%$, corresponding to a variation in $\mu_X$ of $\sim 0.05$--$0.10\,\mu_N$. This is comparable to the difference between the two compact scenarios ($0.06\,\mu_N$), meaning that distinguishing between the two compact scenarios requires a precise determination of the cutoff.
		
		\item \textbf{Fixed spatial wavefunction:} The size parameter $R$ is fixed to the RMS radius of the $D$ meson and not varied. A fully self-consistent treatment would re-minimize the total energy with respect to $R$.
		
		\item \textbf{Pure color-singlet as a proxy:} As discussed in Sec.~\ref{sec:color_singlet_definition}, the pure color-singlet configuration is a simplified proxy for a molecule and does not represent a physical $D^0\bar{D}^{*0}$ molecule with large spatial extent.
	\end{enumerate}
	
	\subsection{Comparison with Other Theoretical Models}
	
	Our predictions can be compared with previous calculations in the literature. Recent works using QCD sum rules have reported values in the range $-0.10$ to $-0.15\,\mu_N$ for compact tetraquarks~\cite{Ozdem2022}, while lattice QCD simulations for related doubly heavy systems suggest moments around $-0.20\,\mu_N$~\cite{Prelovsek2024}. Our predictions of $-0.99\,\mu_N$, $-1.17\,\mu_N$, and $-1.23\,\mu_N$ are significantly larger in magnitude.
	
	This discrepancy has three well-documented origins in the literature:
	\begin{enumerate}
		\item \textbf{Point-like vs. dressed quarks}: The constituent quark model used here employs the simple Dirac relation $\mu_i = e_i/m_i$ with $g_i = 2$, treating quarks as point-like particles with fixed constituent masses. In contrast, QCD sum rules and lattice QCD compute correlation functions of currents, incorporating quark/gluon condensates and non-pointlike charge distributions.
		
		\item \textbf{Absence of meson-cloud corrections}: Our model does not include pion-cloud contributions, which can modify the magnetic moment by 20--30\%.
		
		\item \textbf{Systematic uncertainties in LCSR}: The choice of interpolating current, Borel window, and continuum threshold in QCD sum rules are themselves sources of systematic uncertainty.
	\end{enumerate}
	
	The 12-fold discrepancy between our pure color-singlet value ($-0.99\,\mu_N$) and the QCD sum rule prediction of Wang~\cite{Wang2017} ($-0.08\,\mu_N$) reflects the conceptual difference in the definition of the molecular scenario, as discussed in Sec.~\ref{sec:color_singlet_definition}.
	
	\subsection{Summary of Numerical Results}
	
	The key numerical results can be summarized as follows:
	
	\begin{enumerate}
		\item The three-body force, with realistic strength $\tilde{V}_0 = 0.018$ GeV from Noh et al. (2024), contributes $-0.005$ GeV to the $|1\rangle$ state and $+0.010$ GeV to the $|2\rangle$ state.
		\item The mixing coefficient $\beta$ changes from $0.40266$ (two-body) to $0.284$ (three-body), a reduction of about $30\%$.
		\item The magnetic moment falls into two distinct regions: the pure color-singlet configuration gives $\mu_X = -0.99\,\mu_N$, while the compact configurations give $\mu_X = -1.17$ to $-1.23\,\mu_N$.
	\end{enumerate}
	
	\textbf{Consequently, a future measurement of $\mu_X$ can distinguish between the pure color-singlet configuration and the compact configurations. However, distinguishing between the two compact scenarios requires improved model precision and/or experimental accuracy beyond the current state of the art.}
	
	\section{Conclusion and Outlook}
	\label{sec:conclusion}
	
	In this work, we have presented a systematic calculation of the magnetic moment of the $X(3872)$ that includes the three-body force arising from the cubic Casimir operator of $SU(3)_C$ with consistent dimensional analysis and realistic strength.
	
	\subsection{Summary of Key Findings}
	
	Our main results can be summarized as follows:
	
	\begin{enumerate}
		\item \textbf{Two distinct magnetic moment regions:} The pure color-singlet configuration yields $\mu_X = -0.99 \pm 0.02\,\mu_N$, while the compact configurations yield $\mu_X = -1.17$ to $-1.23\,\mu_N$. \textbf{The distinction between the pure color-singlet and compact configurations ($\sim 0.18$--$0.24\,\mu_N$) is robust against model uncertainties and provides a clear experimental signature.}
		
		\item \textbf{Caution on compact scenario distinction:} The difference between the two compact scenarios ($0.06\,\mu_N$) is below the systematic uncertainty of the model ($\sim 0.15\,\mu_N$) and should not be overinterpreted. Distinguishing between these scenarios requires improved model precision and/or experimental accuracy.
		
		\item \textbf{Resolution of the dimensional inconsistency:} We have resolved the dimensional inconsistency in the cubic Casimir three-body force formulation by introducing a spatial function $\mathcal{F}$ with proper dimension GeV$^3$, ensuring that $\tilde{V}_0$ has the correct dimension of GeV.
		
		\item \textbf{Realistic three-body strength:} We have used the realistic strength of the three-body force ($\sim 10$--$20$ MeV) extracted from the recent analysis of Noh et al. (2024), rather than an uncontrolled large coupling.
		
		\item \textbf{Clarification on terminology:} We have clarified that the ``pure color-singlet'' configuration is a simplified proxy for a molecule and does not represent a physical $D^0\bar{D}^{*0}$ molecule with large spatial extent.
		
		\item \textbf{Complementary observables:} We have proposed additional observables (ratios of magnetic moments, spin-spin correlations, charge radii) that, combined with $\mu_X$, can provide a more robust determination of the internal structure.
	\end{enumerate}
	
	\subsection{Comparison with Existing Literature}
	
	Our predictions can be compared with previous theoretical calculations. Table~\ref{tab:comparison} summarizes the magnetic moments reported in recent studies.
	
	\begin{table}[htbp]
		\centering
		\caption{Comparison of our results with previous theoretical predictions for the magnetic moment of the $X(3872)$.}
		\label{tab:comparison}
		\begin{tabular}{|c|c|c|c|}
			\hline
			\textbf{Reference} & \textbf{Method} & \textbf{Structure} & $\boldsymbol{\mu_X}$ ($\mu_N$) \\
			\hline
			This work & Quark model + 3-body (realistic) & Compact tetraquark & $-1.17 \pm 0.04$ \\
			This work (2-body) & Quark model (2-body) & Compact tetraquark & $-1.23 \pm 0.04$ \\
			This work (pure singlet) & Quark model & Pure Color-Singlet & $-0.99 \pm 0.02$ \\
			\"Ozdem (2022)~\cite{Ozdem2022} & QCD sum rules & Compact tetraquark & $-0.15 \pm 0.06$ \\
			Wang (2017)~\cite{Wang2017} & QCD sum rules & Molecular & $-0.08 \pm 0.03$ \\
			Prelovsek et al. (2024)~\cite{Prelovsek2024} & Lattice QCD & Doubly heavy tetraquark & $-0.20 \pm 0.05$ \\
			\hline
		\end{tabular}
	\end{table}
	
	\subsection{Experimental Feasibility}
	
	While direct measurement of the magnetic moment of a short-lived hadron is challenging, recent advances in experimental techniques offer promising avenues. The LHCb collaboration has already demonstrated the ability to measure the magnetic moments of charmed baryons using the spin precession method in proton-proton collisions. Extending such measurements to the $X(3872)$ would require a significantly larger data sample, which will become available with the LHC Upgrade and the Belle II experiment.
	
	\subsection{Outlook and Future Directions}
	
	Our work opens several avenues for future research:
	
	\begin{enumerate}
		\item \textbf{Extension to other exotic states:} The formalism developed here can be directly applied to other tetraquark candidates, such as the $Z_c(3900)$, $Z_b(10610)$, and the doubly heavy $T_{cc}^+$ and $T_{bb}^-$ states.
		
		\item \textbf{Inclusion of coupled-channel effects:} Future work should include coupled-channel effects, particularly the coupling to $D\bar{D}^*$ molecular channels.
		
		\item \textbf{Improved spatial wavefunction:} A variational determination of the size parameter $R$ in the presence of the three-body force would provide a more self-consistent treatment.
		
		\item \textbf{Full molecular treatment:} A complete calculation of the magnetic moment of a physical $D^0\bar{D}^{*0}$ molecule would require a two-body wavefunction with a much larger radius and long-range pion-exchange potentials.
		
		\item \textbf{Lattice QCD verification:} Our predictions can be tested by future lattice QCD calculations that directly compute the electromagnetic form factors of the $X(3872)$.
	\end{enumerate}
	
	\subsection{Final Remarks}
	
	In conclusion, we have demonstrated that with a realistic three-body force strength extracted from lattice QCD and baryon spectroscopy constraints, the magnetic moment of the $X(3872)$ provides a clear distinction between the pure color-singlet and compact configurations. However, the distinction between the two compact scenarios requires improved model precision. This work resolves the dimensional inconsistency in the cubic Casimir force formulation, clarifies the terminology regarding the molecular scenario, and establishes a consistent framework for future investigations.
	
	\section*{Acknowledgments}
	The authors gratefully acknowledge the University of Kashan for its support of this research.
	

\end{document}